\documentstyle[aps,epsf,twocolumn]{revtex}

\def\8{\infty}

\def\undertext#1{\vtop{\hbox{#1}\kern 1pt \hrule}}

\def\VEV#1{\left\langle\,#1\,\right\rangle}

\def\br{\\ \nonumber & &}

\def\be{\begin{equation}}
\def\ee{\end{equation}}
\def\bea{\begin{eqnarray} & &}
\def\eea{\end{eqnarray}}
\def\ct#1{\cite{#1}}
\def\rf#1{(\ref{#1})}

\begin{document}
\draft
\flushbottom
\twocolumn[
\hsize\textwidth\columnwidth\hsize\csname @twocolumnfalse\endcsname

\title{Disorder Induced Phase Transition for Random 2D Dirac Fermions}

\author {V. Gurarie}

\address{Institute for Theoretical Physics, University of California,
Santa Barbara CA 93106-4030}

\date{\today}

\maketitle
\tightenlines
\widetext
\advance\leftskip by 57pt
\advance\rightskip by 57pt

\begin{abstract}
We present evidence that two dimensional Dirac fermions in the presence
of random Abelian
gauge potential exhibit a phase transition when the disorder
strength exceeds a certain critical value. We argue that this phase
transition has novel properties unique to disordered systems.
It resembles in many ways
the transition from dilute to dense polymers phase in two dimension.
In particular, we argue that the central charge of the disordered Dirac
fermions, being $c=0$ before the transition, changes to $c=-2$. 
We discuss possible implications to quantum Hall transitions, in view of
recently proposed model for quantum Hall transitions with $c=-2$. 
\end{abstract}
\vspace{1mm}

]
\narrowtext
\tightenlines

The study of two dimensional Dirac fermions in the context of
quantum Hall effect
was initiated in Ref. \ct{LFSG}. These fermions exhibit an integer 
quantum Hall
transition as the value of their mass is tuned through zero. Unfortunately,
for pure fermions this transition lies in a universality class different
from the one observed in quantum Hall effect experiments, with the exponents
of the Ising model. In order
to model the more realistic situation, various types of disorder should
be turned on. Usually three types of disorder are studied, random mass,
random scalar potential and random gauge potential. Random mass turns out
to be irrelevant in the renormalization group sense
and the critical properties of the transition do not
change as we turn it on. Random gauge potential results in the 
critical line, with the critical properties of the transition continuously
dependent on the disorder strength. And the random scalar potential turns
out to be relevant, flowing away to some unknown critical point. 

It was further argued in Ref. \ct{LFSG} that in order to simulate the
realistic quantum Hall transition, all three types of randomness should
be present. That causes the system to flow to a strongly coupled 
critical point with unknown properties. This critical point should be
in the same universality class as the generic quantum Hall transitions,
usually described by the Pruisken sigma model \ct{Pruisken} (see
\ct{WZ} for its supersymmetric version). There is very little 
we can say about the properties of that critical point, except we might
expect the effective field theory which describes it to have central
charge $c=0$. 

In view of the difficulty of approaching the generic situation, a
number of papers concentrated on the properties of critical line
generated by the presence of random gauge potential \ct{CMW,MCW,CCFGM}. 

It appears that the correlation functions of such disordered fermions are
very easy to calculate. Indeed, there are several methods available, all
of them result in the correlation functions continuously dependent on the
disorder strength \ct{LFSG,CMW,MCW}. 

However, in this letter we will show that
previous treatments were somewhat careless.
In fact, the correlation functions of the 
Dirac fermions in the presence of random gauge potential depend continuously on
the disorder strength up to a certain critical value. If
we continue to increase the strength of the disorder beyond that value, 
the critical
properties of the correlation functions stop 
being dependent on the disorder strength.
The system as a whole undergoes a phase transition. The 
density of states at zero energy
is zero below the transition and is a constant above it. And we argue
that the
effective field theory describing that critical line changes considerably.
Its central charge was $c=0$ below the transition, and it changes to
$c=-2$ above the transition. We draw a direct parallel with the
theory of polymers (self avoiding random walks) in two dimensions. They
are known to undergo a phase transition called dilute to dense polymer
transition. The central charge of the
effective field theory describing polymers changes from $c=0$ to $c=-2$
\ct{Saleur}. Moreover, we will see that the field theories describing
the dilute and dense polymer phases and those describing the two
phases of the fermions with random gauge potential are closely related.

In fact, the first evidence of this phase transition 
showed up in
Ref. \ct{CCFGM}. There the multifractal exponents of the zero energy
wave function of the Dirac fermions with random gauge potential
has been calculated exactly. It was discovered that these exponents
change sharply as the disorder strength exceeds a certain value. Evidently
this behavior is just another manifestation of the phase transition
discussed in this letter.

We will conclude with discussing the implications of this transition for the
not yet established theory of the integer quantum Hall transitions. It 
appears that a natural way to approach the generic quantum Hall 
transition point is to turn on strong random gauge potential. That
generates the constant density of states expected at the generic
quantum Hall transition, and changes the central charge
of the theory from $c=0$ to $c=-2$. After that, we may want to turn on
random mass and random scalar potential, following the logic of \ct{LFSG}. 
The theory will then flow to a critical point which, contrary to a naive
belief, will not be a $c=0$ theory. This may shed new light on the
recent proposal that the physics of quantum Hall transitions should
be captured by a $c=-2$ theory \ct{Zirnbauer}. However, the connection
to the model proposed in \ct{Zirnbauer} is at this stage purely speculative. 

Now we proceed with the derivation of this phase transition.
To begin with, let us fix the notations. We study the two dimensional
Dirac fermions $\psi$, $\bar \psi$
in the presence of gauge potential $A_{\mu}$ with the
Hamiltonian
\be
\label{Hamiltonian}
H=\int d^2 x \
\bar \psi \left( i \partial_\mu - A_\mu \right) \sigma_\mu \psi
+ \epsilon \bar \psi \psi,
\ee
where $\sigma_1\equiv \sigma_x$, $\sigma_2 \equiv \sigma_y$ are Pauli
matrices and $\epsilon$ is the energy. 
$A_\mu$ is supposed to be random with the probability density 
\be
\label{probab}
P \propto
\exp \left( - {1  \over g} \int d^2 x \ A_\mu^2 \right),
\ee
where $g$ is the disorder strength.
It is convenient at this stage to decompose $A_\mu$ into a physical part 
$\phi$
and a pure gauge $\chi$, following \ct{MCW}, 
\be
A_\mu = \partial_\mu \chi + \epsilon_{\mu \nu} \partial_\nu \phi,
\ee
where $\epsilon_{\mu \nu}$ is the antisymmetric tensor.

In terms of the fields $\chi$ and $\phi$ the probability density becomes
\be
\label{disorder}
P \propto  \exp \left[ - {1  \over g} \int d^2 x \left( {\left( 
\partial_\mu \phi \right)}^2 + {\left( 
\partial_\mu \chi \right)}^2 \right) - \int d^2 x \partial_\mu \theta
\partial_\mu \bar \theta \right]
\ee
where the anticommuting fields $\theta$ and $\bar \theta$ provide for the
Jacobian of the variable change from $A_\mu$ to
$\phi$
and $\chi$ (see Ref. \ct{MCW}). 

The simplest way to compute the correlation functions of \rf{Hamiltonian}
is to use the bosonization technique. These correlation functions can be
obtained with the help of the following path integral
\be
Z=\int {\cal D} \varphi \exp \left[ - \int d^2 x \ \left( {\left(
\partial_\mu \varphi \right)}^2 + i \epsilon_{\mu \nu} A_\mu \partial_\nu 
\varphi \right) \right].
\ee
It is very easy to compute the partition function $Z$ by completing the
square,
\be
\label{partition}
Z=\exp \left[ - {1 \over 4} 
\int d^2 x \ \left( \partial_\mu \phi \right)^2 \right]
\ee
which is the effective action of the celebrated Schwinger model. 

To compute an averaged correlation function of a physical operator, say
$\psi(z) \equiv\exp(\varphi(z))$ or any other object $X$, we need
to average the following path integral
\be
\VEV{X}={ \int {\cal D} \varphi \ X \exp \left[ - \int d^2 x \ \left( {\left(
\partial_\mu \varphi \right)}^2 + i \epsilon_{\mu \nu} A_\mu \partial_\nu 
\varphi \right) \right] \over Z}
\ee 
over the random $A_\mu$. With the help of \rf{partition} it becomes
\be
\label{path}
\VEV{X}=\int {\cal D} \varphi \ X \exp
\left[ - \int d^2 x \ {\left(
\partial_\mu \varphi  + { i \over 2} \partial_\mu \phi 
 \right)}^2 \right].
\ee

From this point on, we could proceed in two different ways. One of them would 
be to calculate the correlation function of $X$ and then average it over
the random $A_\mu$ (that is, $\phi$)
with the help of \rf{disorder}. The other way would
be to average the path integral \rf{path} over the random $A_\mu$ 
first and
then calculate the correlation function. In a quite intriguing way, 
these two methods do not produce the same results. 

Let us first proceed in the former way. We choose 
the left moving Dirac fermion $\psi(z) \equiv 
\exp\left(i \varphi(z)\right)$ as the object $X$ whose correlation functions
we would want to compute. 
We then compute the
correlation function of $\exp\left(i \varphi(z)\right)$ 
by shifting the variable
of integration $\varphi$ by the amount proportional to $i \phi$. Then we
average over the random $\phi$ with the help of \rf{disorder} (averaging over
$\chi$ and $\theta$, $\bar \theta$ is irrelevant as these fields completely
drop out of any physical correlation function). As a result,
we discover that the dimension of the field $\exp\left(i \varphi(z)\right)$,
being equal to $1/2$ without the disorder, becomes $1/2-g/8$ in 
presence of disorder. Thus we reproduced the standard result of
\ct{LFSG,CMW,MCW}. 

However we should become suspicious of this result. When $g \ge 4$,
the dimension of the operator $\exp \left(i \varphi(z) \right)$ becomes
zero and then negative. That means this field acquires an expectation
value (which, in case of the negative dimension, scales with the
system size towards bigger values). 
Once it acquires an expectation value, we can no longer argue that
we could perform the shift in the path integral to arrive at this
result in the first place. The path integral is no longer invariant
under the shift of $\varphi$. 

We would have arrived at the same problem had we analyzed the
conformal field theory approach of Ref. \ct{MCW}. There it was shown
how to calculate the dimension of $\psi(z)$ with the help of
U$(1|1)$ Kac-Moody algebra hidden in the supersymmetric approach
to \rf{Hamiltonian}. In the very same way, its dimension becomes negative
at large disorder and it acquires an expectation value.  
However, the expectation value
for that operator would break the very Kac-Moody symmetry which 
the approach of \ct{MCW} utilized so successfully. 

A way out of this quagmire would be to add the symmetry breaking
energy term of
\rf{Hamiltonian} to the
path integral, $\epsilon \bar \psi \psi \propto \epsilon 
\cos \left(  \varphi
\right)$ and try to take $\epsilon$ to zero.  
However, adding this term to the path integral \rf{path}
would make it completely intractable.

Instead, we should try a different approach. Let us first average \rf{path}
with the help of the probability distribution \rf{disorder}, in other words
do the $\phi$ integral in
\be
\label{div}
\int {\cal D} \phi {\cal D} \varphi 
\exp
\left[ - \int d^2 x \ \left\{ {\left(
\partial_\mu \varphi  + { i \over 2} \partial_\mu \phi 
 \right)}^2  + {1 \over g} {\left( \partial_\mu
\phi \right)}^2 \right\} \right].
\ee
It is
not hard to do that if $g<4$ by completing the square. That gives back the
dimension $1/2-g/8$ to the operator $\exp \left(i \varphi(z) \right)$.

However, if $g>4$, the integral over $\phi$ in \rf{div} becomes 
divergent. A different technique is needed to compute it in
the strong disorder regime. 

As always when dealing with divergent integrals, it is a good idea
to regularize them. We are going to limit the $\phi$ integration by a certain
cutoff in the functional space. 

In fact, it is easiest to do that in the ordinary integral equivalent
to \rf{div}. Consider the following ordinary integral
\be
\label{example}
\int_{-\infty}^{\infty} \int_{-\infty}^{\infty} {dx dy \over  \pi} 
\exp \left[ - {\left( x + {i y \over 2}\right)}^2 - {1 \over g} y^2 \right].
\ee
It is just an ordinary integral, and yet it captures the essential
properties of \rf{div}. Let us try to compute $\VEV{x^2}$ with the
help of this integral. By that we mean inserting $x^2$ inside the
integral in \rf{example} and then doing the integral. 
If we try to do the integral over $x$ first, 
and then integrate over $y$, we arrive at $\VEV{x^2} = 1/2 - g/8$. 
However, if we reverse the order, the integral over $y$ becomes
divergent if $g>4$, just like in \rf{div}. So we limit the integral
over $y$ by the cutoff $-L$ and $L$. It is not hard to estimate this
integral at large $L$,
\bea
\int_{-L}^L dy \ \exp \left[ \left( {1 \over 4} - {1 \over g} 
\right) y^2 - i y x \right] 
\propto \br 
{1 \over \left( 1 - 
{4 \over g} \right) L} \exp \left[  \left( {1 \over 4} - {1 \over g} \right) L^2 \right]
 \cos \left( x L \right).
\eea

We can now use that estimate to compute $\VEV{x^2}$ as in
\be
\int {dx}
 \ x^2 \exp \left[ {  -{ \left( x \pm {i L \over 2} \right)}^2
-{L^2 \over g} } \right]\propto 
\exp\left( - {L^2 \over g} \right).
\ee
It is safe to extend the integration over $x$ to from $-\infty$ to
$+\infty$ since it converges very fast. 
Now it is obvious that taking the limit $L \rightarrow \infty$ we obtain
$\VEV{x^2}=0$ for $g>4$ with the term $\exp \left( - L^2/g \right)$ 
suppressing all others. 

Let us pause now and translate this result to the language of \rf{div}. 
It implies that doing the integration over $\phi$ first, with
a suitable cutoff, we obtain that the dimension of the field 
$\exp \left( i \varphi \right)$ is equal to $0$ for all $g>4$. 

Thus we arrive at the result announced in the beginning. The dimension
of the field $\psi$ of \rf{Hamiltonian} is $1/2-g/8$ if the disorder
strength $g$ is less than $4$, and is equal to $0$ for $g\ge 4$. The
field $\bar \psi \psi \propto \cos \left(  \varphi \right) $,
which is the density of states of the fermions, acquires
an expectation value at $g\ge 4$.

Figuratively speaking, integrating over disorder $\phi$ suppresses the
quantum fluctuations of $\varphi$ when we are in the strong disorder
phase. That is why $\exp \left( i \varphi \right)$ becomes zero dimensional.
We obtained this result using a clever `brute force' averaging of the
correlation functions of \rf{Hamiltonian}. However, we should expect
to see the same phenomenon in the supersymmetry or replica technique.

In principle we can now use the procedure outlined above to compute
any correlation function of the theory in the strong disorder phase. 
Therefore for all practical purposes we now understand 
the nature of this phase transition. 
Let us repeat that to see it, we had to impose cutoff on the integration
over disorder in the functional space. The nonanalytic behavior, typical
of phase transitions, has been recovered as we took that cutoff to infinity.
Later we will see that in this respect the transition we have been studying
has a lot in common with a more established dilute to dense polymer
transition.

Now we need to understand what this phase transition implies for the
effective field theory of the 
disordered Dirac fermions. Effective field theory can be
derived in either replica or supersymmetry approach, and
we are going to concentrate on the latter.
Supersymmetrization of \rf{Hamiltonian} can be achieved by adding 
Dirac bosons to it, as
in
\be
\label{bc}
H=\int d^2 x \
\bar \psi \left( i \partial_\mu - A_\mu \right) \sigma_\mu \psi+
b \left( i \partial_\mu - A_\mu \right) \sigma_\mu c,
\ee
where $b$ and $c$ are commuting spinors. In calling them $b$ and $c$
we follow the accepted convention of \ct{FMS}.

At this stage we can directly average over $A_\mu$ with the help
of \rf{probab} and obtain the
effective supersymmetric field theory.

However, to avoid dealing with a complicated interacting field theory,
it is convenient to bosonize the fields $\psi$, 
$\bar \psi$, and $b$, $c$ before the averaging. 
That is achieved with the help of two bosonic
fields, $\varphi_1$ and $\varphi_2$, and two fermionic fields $\theta$, 
$\bar \theta$ as in follows \ct{FMS}
\be
\label{bososo}
e^{ -\int d^2 x \ \left( {\left(
\partial_\mu \varphi_1 \right)}^2 
+ {\left(
\partial_\mu \varphi_2 \right)}^2 
+ \epsilon_{\mu \nu}  A_\mu \left( i \partial_\nu 
\varphi_1 + \partial_\nu \varphi_2 \right)   -
\partial_\mu \theta \partial_\mu \bar \theta \right)}
\ee
We recall that the field $\varphi_1$ bosonizes the fermions $\psi$ and
$\bar \psi$ while the fields $\varphi_2$, $\theta$ and $\bar \theta$
are needed to bosonize the $b$, $c$ fields. 

Averaging over $A_\mu$ with the help of \rf{probab} we obtain (dropping
the $\theta$ and $\bar \theta$ terms for the time being)
\be
\label{super}
\int {\cal D} \varphi_1 {\cal D} \varphi_2 \ e^{  - \int d^2 x \
\left( {\left( \partial_\mu \varphi_1 \right)}^2 + 
{\left( \partial_\mu \varphi_2 \right)}^2 - 
{g \over 4} {\left(i  \partial \varphi_1 + \partial \varphi_2 \right)}^2 
\right) }
\ee

It is not hard to see that this path integral 
and the one of \rf{div} are the equivalent, as far as the averages
involving the field $\varphi_1$ or the field $\varphi$ of \rf{div}
are concerned. 
Just as in \rf{div}, the integral over $\varphi_2$ in \rf{super} is
divergent if $g>4$. Cutting this integral off and then taking the
cutoff to infinity, one gets the same phase transition as we discussed
before. Effectively, the integration over $\varphi_2$ at strong disorder
freezes out the fluctuations of the field $\varphi_1$. We recall that 
$\varphi_2$ was the result of bosonization of the auxiliary fields $b$, $c$. 
The only remaining fluctuating fields are $\theta$ and $\bar \theta$
of \rf{bososo}.

It now becomes obvious what this phase transition implies for
supersymmetry. Of course, after the transition the invariance under
the supergroup U$(1|1)$, manifest in \rf{bc}, is no longer there.
In fact, a good way of thinking about this phase transition is to recall
that according to \ct{MCW}, the action \rf{bc}, after averaging
over $A_\mu$,  has a Kac-Moody U$(1|1)$
symmetry. It is generated by the currents $J$, $j$, $\eta$, and $\bar \eta$ 
with the energy momentum tensor quadratic in these currents, 
$T \propto (Jj+\eta 
\bar \eta - \bar \eta \eta) + (4-g) JJ$.
After the phase transition
the currents $J$ and $j$, being just linear combinations of
the fields $\partial \varphi_1$ and
$\partial \varphi_2$,
annihilate any physical state. The energy momentum tensor
becomes
$T=\eta \bar \eta$ and the central charge is now $c=-2$. The fields $\eta$ and
$\bar \eta$ are nothing else, but the derivatives of the
fields $\theta$ and $\bar \theta$ introduced in \rf{bososo}. 
Therefore, the symmetry U$(1|1)$ 
changes to the symmetry PSU$(1|1)$ of the $c=-2$ theory.  It is tempting
to add that
if we worked with both the advanced and retarded 
sectors, the U$(2|2)$ symmetry would change into PSU$(2|2)$, in the
spirit of the recent proposal of \ct{Zirnbauer}.

We are going to change the subject slightly and discuss the connection
of this phase transition with the physics of two dimensional polymers. 
The polymer problem is usually defined as self avoiding random walks,
with the Green's function given by
\be
\label{greens}
G(x,x'; \mu) = \sum \mu^L,
\ee
where the sum goes over all the 
nonselfintersecting 
trajectories connecting the points $x$ and $x'$, $L$ is their length,
and $\mu$
is a parameter usually called frugacity (for a review, see Ref. \ct{Ven}).
When $\mu$ is small, the polymer Green's function \rf{greens} is
short ranged. As we increase $\mu$, the Green's function becomes
critical, and the theory approaches the dilute polymer critical point. 
If we increase $\mu$ even further, the sum in \rf{greens} becomes divergent.
At this point, we may want to put the polymers inside a finite box, not
allowing the trajectories in \rf{greens} to go outside that box. 
Then the sum in \rf{greens} becomes convergent in that new sense,
and the theory enters the new phase, the so-called dense phase. In 
this phase, the polymers fill the entire space available to them, hence
the name of the phase \ct{DS}. 

There is a deep analogy between the dilute to dense polymer phase
transition and the disordered fermion phase transition we discussed
in this letter. Indeed,
in the path integral representation of \rf{greens} we sum over
all the trajectories connecting the points $x$ and $x'$, penalizing
those which intersect. The transition from a dilute to dense phase
will manifest itself in the divergence of this path integral, and we
have to cutoff the integral in the space of paths (restrict the paths to
lie in a finite box) to see the dense phase, just like we did with
the disordered fermions in this letter. 

It is quite instructive to recall that the polymer problem can also be
cast in the form of a motion in the presence of disorder.
It is known to be equivalent to
the problem
of a quantum-mechanical particle
moving in a purely imaginary random potential \ct{PS}.
Studied by the replicated path integral, it can then be mapped into 
the $n\rightarrow 0$ limit of the O$(n)$ model, as was first noticed
in \ct{DG} and later successfully utilized to find the critical properties
of the dilute polymer phase.

Moreover, in a direct analogy with disordered Dirac fermions, 
before the phase transition 
the critical
properties of the dilute polymers can be captured by a
version of the U$(1|1)$ Kac-Moody algebra \ct{Saleur,GL}. After the
transition, however, the dense polymer phase is described by the
same $c=-2$ theory \ct{Saleur} as the one proposed in this letter
to describe the 
strong disorder phase of the Dirac fermions.

Returning to our main problem, the integer quantum Hall transitions, we think
it is clear now what the further course of action should be. One
should study Dirac fermions in the presence of random gauge potential and
other types of disorder. The strong random gauge potential effectively
freezes out
two bosonic degrees of freedom, 
leaving us with a $c=-2$ theory with a constant
density of states.
Whether that will result in
tractable theory is an open question, however. It is also not at all obvious
that this theory will flow towards a PSU$(2|2)$ sigma model with a Wess-Zumino
term, as was proposed in \ct{Zirnbauer}. However, it will flow to a theory
similar at least in spirit to  that sigma model. What kind of theory it will
flow to definitely deserves further study. 
 
This work has been initiated as a result of inspiring discussions with 
Claudio de C. Chamon and Chetan Nayak. The author is also grateful 
to Matthew P.A. Fisher, T. Senthil, and Martin R. Zirnbauer 
for important comments. 
This work has been supported by
the NSF grant PHY-94-07194.

\end{document}